\documentclass[letterpaper]{iopconfs}

\usepackage{graphicx}
\usepackage{color}
\usepackage{amssymb}

\begin{document}

\title{Progress towards fabrication of $^{229}$Th-doped high energy band-gap crystals for use as a solid-state optical frequency reference}

\author{Wade~G.~Rellergert$^1$, Scott~T.~Sullivan$^1$, D.~DeMille$^2$, R.~R.~Greco$^3$, M.~P.~Hehlen$^3$, R.~A.~Jackson$^4$, J.~R.~Torgerson$^3$, and Eric~R.~Hudson$^1$}
\address{$^1$Department of Physics and Astronomy, University of California, Los Angeles, California 90095, USA \newline $^2$Department of Physics, Yale University, New Haven, Connecticut 06511, USA \newline $^3$Los Alamos National Laboratory, Los Alamos, New Mexico 87545, USA \newline $^4$School of Physical and Geographical Sciences, Keele University, Keele, Staffordshire ST5 5BG, UK}

\ead{eric.hudson@ucla.edu}

\begin{abstract}
We have recently described a novel method for the construction of a solid-state optical frequency reference based on doping $^{229}$Th into high energy band-gap crystals~\cite{rellergert_constraining_2010}. Since nuclear transitions are far less sensitive to environmental conditions than atomic transitions, we have argued that the $^{229}$Th optical nuclear transition may be driven inside a host crystal resulting in an optical frequency reference with a short-term stability of $3\times10^{-17}<\Delta f/f <1\times10^{-15}$ at 1~s and a systematic-limited repeatability of $\Delta f/f \sim 2 \times 10^{-16}$. Improvement by $10^2-10^3$ of the constraints on the variability of several important fundamental constants also appears possible. Here we present the results of the first phase of these experiments. Specifically, we have evaluated several high energy band-gap crystals (Th:NaYF, Th:YLF, Th:LiCAF, Na$_2$ThF$_6$, LiSAF) for their suitability as a crystal host by a combination of electron beam microprobe measurements, Rutherford Backscattering, and synchrotron excitation/fluorescence measurements. These measurements have shown LiCAF to be the most promising host crystal, and using a $^{232}$Th doped LiCAF crystal, we have performed a mock run of the actual experiment that will be used to search for the isomeric transition in $^{229}$Th.  This data indicates that a measurement of the transition energy with a signal to noise ratio (SNR) greater than 30:1 can be achieved at the lowest expected fluorescence rate.
\end{abstract}


\section{Introduction}

The existence of a low-lying state in $^{229}$Th, with an estimated natural linewidth, $\Gamma_n$, in the range $0.1 \leq \Gamma_n/2 \pi \leq  10$~mHz~\cite{tkalya_decay_2000, ruchowska_nuclear_2006}, was indirectly established over 3 decades ago~\cite{kroger_features_1976}.  However, direct searches for emission from this M1 transition have so far been unsuccessful~\cite{tkalya_decay_2000}.
Recently, a new indirect measurement has established that the transition energy is 7.8~$\pm$~0.5~eV~\cite{beck_energy_2007}, rather than 3.5~eV as orginally thought.  Although well into the vacuum ultraviolet (VUV), this energy is amenable to study by laser spectroscopy, opening the door for use of this transition as a highly stable oscillator in next generation optical clocks.  The transition is also predicted to show a 10$^6$-fold enhancement in sensitivity to the time variation of several fundamental constants\cite{flambaum_enhanced_2006, berengut_proposed_2009}, \textit{e.g.} the fine structure constant.  While the current measurement uncertainty of the nuclear transition energy is very good by nuclear physics standards, it must be drastically improved in order to realize the exciting applications of this transition.  

We have recently proposed a novel approach to \textit{directly} measure the transition energy to a much higher precision by interrogating $^{229}$Th that has been doped into a VUV transparent crystal~\cite{hudson_investigation_2008, peik_prospects_2009, rellergert_constraining_2010}.  Using this approach more than 10$^{19}$ Th nuclei can be interrogated, which allows for fluorescence rates sufficient to directly measure the transition energy using a broadband synchrotron light source.

In this article, we present measurements of inherent VUV-induced background signals which could mask fluorescence from the isomer state in $^{229}$Th for several candidate host crystals. While the data shows that many of the crystals display some spurious fluorescence, the fluorescence is in general of a low level and short-lived compared to that expected from the isomeric transition.  Of all the crystals measured, we have determined LiCAF (LiCaAlF$_6$) to be the most suitable as it yields very little background fluorescence when it is directly illuminated with VUV light.  Additonally, it shows resilience to VUV-induced radiation damage and has been grown with $^{232}$Th as a dopant at the 0.05\% level.  It also has several key structural properties that mitigate broadenings of the isomeric transition energy in a crystal environment.

\section{Experimental Results}

The first steps towards realizing the exciting applications of this transition are identifying suitable host crystals and determining the transition frequency to a greater precision. The key requirements for the host crystal are that it be reasonably transparent in the VUV, have a pure crystalline structure, and chemically accept Th$^{4+}$ ions. Numerous candidate hosts can be identified in the extensive family of fluoride crystals. Potential crystals include Na$_2$ThF$_6$, LiCAF (LiCaAlF$_6$), LiSAF (LiSrAlF$_6$), YLF (YLiF$_4$), and CaF$_2$, which can be grown by various techniques, \textit{e.g.} Czochralski, Bridgman, etc. ~\cite{Burkhalter2001, Samtleben2005}. While these crystals have band-gaps of $>$10\,eV, impurities and color centers can cause residual absorption and may interfere with the $^{229}$Th optical spectroscopy. Various metal ions as well as oxygen-based impurities at the part-per-million level can lead to absorption coefficients of up to 0.1\,cm$^{-1}$~\cite{Sabatini1975}.  To ensure crystal transparency,  purification methods must be employed to reduce these impurities to part-per-billion levels~\cite{Hehlen2009}. We have applied these methods to the preparation of high purity metal fluoride precursors for crystal growth.

So far, we have studied crystals that were obtained commercially from AC Materials.  These include Na$_2$ThF$_6$, Th:LiCAF, Th:YLF, and Th:NaYF which were grown using the less expensive, but chemically identical, $^{232}$Th isotope.  The Th:NaYF and Th:YLF were grown with a target doping of 0.1\% molar while the Th:LiCAF was grown with a target doping of 1\% molar.  These crystals are good control crystals as the isomeric transition is not present in $^{232}$Th, but are chemically identical and thus the doping process as well as  all VUV induced backgrounds are the same.  They therefore allow us to study which crystals are most suitable for higher precision measurements of the isomeric transition energy. 

\subsection{Synchrotron excitation measurements}

As previously mentioned, a major advantage to our approach for measuring the isomer energy is the large number of $^{229}$Th nuclei that can be addressed.  This allows for fluorescence rates sufficient for the use of standard spectroscopic techniques to determine the transition energy to a higher precision.  We plan to use a multi-phase experimental approach~\cite{rellergert_constraining_2010} to investigate the ulimate precision to which this transition energy can be measured.  

The first phase will be a low resolution study using the 9.0.2.1 beam line at the Advanced Light Source (ALS) at Lawrence Berkeley Labs.  The ALS provides a tunable VUV light source (5~eV-30~eV) with a linewidth ($\Delta$) of 0.175~eV, a photon flux ($\Phi_{p}$) of $\sim$10$^{20}$ photons/cm$^2\cdot$s, and a 170~$\mu$m~$\times$~50~$\mu$m spot size.  Assuming both the light ($\Delta$) and broadened transition ($\Gamma$) linewidths to be Lorentzian, then far from saturation, the excitation rate, $\frac{dN_{e}}{dt}$, is given by
\[\frac{dN_{e}}{dt} \approx  \left( \frac{\lambda}{2 \pi} \right)^2  \frac{\Gamma_n}{\Gamma + \Delta} \frac{1}{1+4(\frac{\omega_0 - \omega_L}{\Gamma + \Delta})^2} \Phi_{p} N_{g}\]
where $\omega_0$ ($\lambda$) is the transition frequency (wavelength), $\omega_L$ is the center frequency  of the ALS beam, and $N_{g}$ is the total number of ground state $^{229}$Th nuclei addressed by the ALS beam.  To locate the transition the ALS center frequency will be scanned across the interesting range of (7.8~$\pm$~0.5~eV) in steps of 0.05~eV with 200~s of  illumination followed by 100~s of light collection at each point.  Even in the lowest fluorescence case ($\Gamma_n = 2 \pi \times 100$~$\mu$Hz), we expect a minimum total fluorescence rate of $\sim$15~kHz when the ALS is on resonance.  

Once the resonance is found, longer illumination will lead to sufficient fluorescence rates for a VUV spectrometer measurement. We expect that the transition frequency can be determined to $\sim$50~cm$^{-1}$ (0.1~nm) in these experiments, but this of course relies on sufficiently low background fluorescence rates.  As a result, we have performed test runs on our control crystals using the VUV light beam at the ALS.

\begin{figure}[h]
\resizebox{0.95\columnwidth}{!}{
    \includegraphics{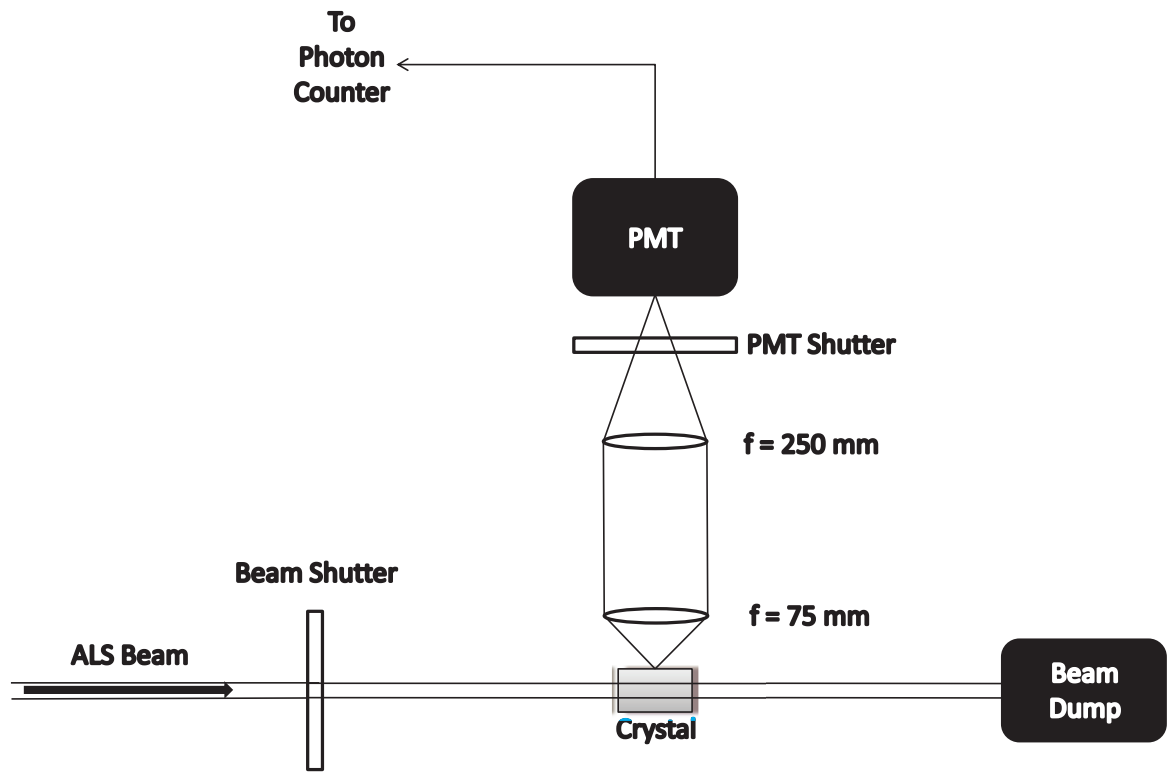}%
} \caption{Figure of the experimental setup.
Fluorescence was collected perpendicular to the ALS beam line via a
2-lens system. The first lens was 75~mm
from the crystal. The fluorescence was imaged onto the
PMT photocathode using a 250~mm lens. \label{ALS_Setup}}
\end{figure}

The setup used is shown in Fig.~\ref{ALS_Setup}.  Each control crystal (3~mm $\times$ 3~mm $\times$ 10~mm) was illuminated along its long axis with the ALS beam and any resultant fluorescence from the crystal was collected in a direction orthogonal to the beam via a 2-lens system. Both lenses were fabricated from LiF and held in vacuum.  The fluorescence was imaged onto the photocathode of a photomultiplier tube (PMT) (Hamamatsu R7639).  The PMT was either under a constant Argon purge or held in vacuum to mitigate VUV absorption by air.

As shown in Fig.~\ref{ALS_Setup}, two pneumatic shutters were mounted inside the vacuum chamber.  One shutter was used to prevent VUV scattered light from hitting the PMT during crystal illumination, and the second shutter was used to block the ALS beam before photon collection with the PMT began.  This introduced a small delay between the point when illumination ended and photon collection began, but in all cases the delay was less than 28~ms.  As the shortest prediction for the $^{229}$Th isomer lifetime is $\sim$100~s this is sufficiently fast to investigate any spurious long-lived background fluorescence signals in the host crystal that might interfere with measuring fluorescence from the isomeric transition.  In what follows, we present the data for all crystals that we have measured to date.

\subsubsection{Th:NaYF}\ 

\begin{figure}[h]
\begin{minipage}{19pc}
\includegraphics[width=21pc]{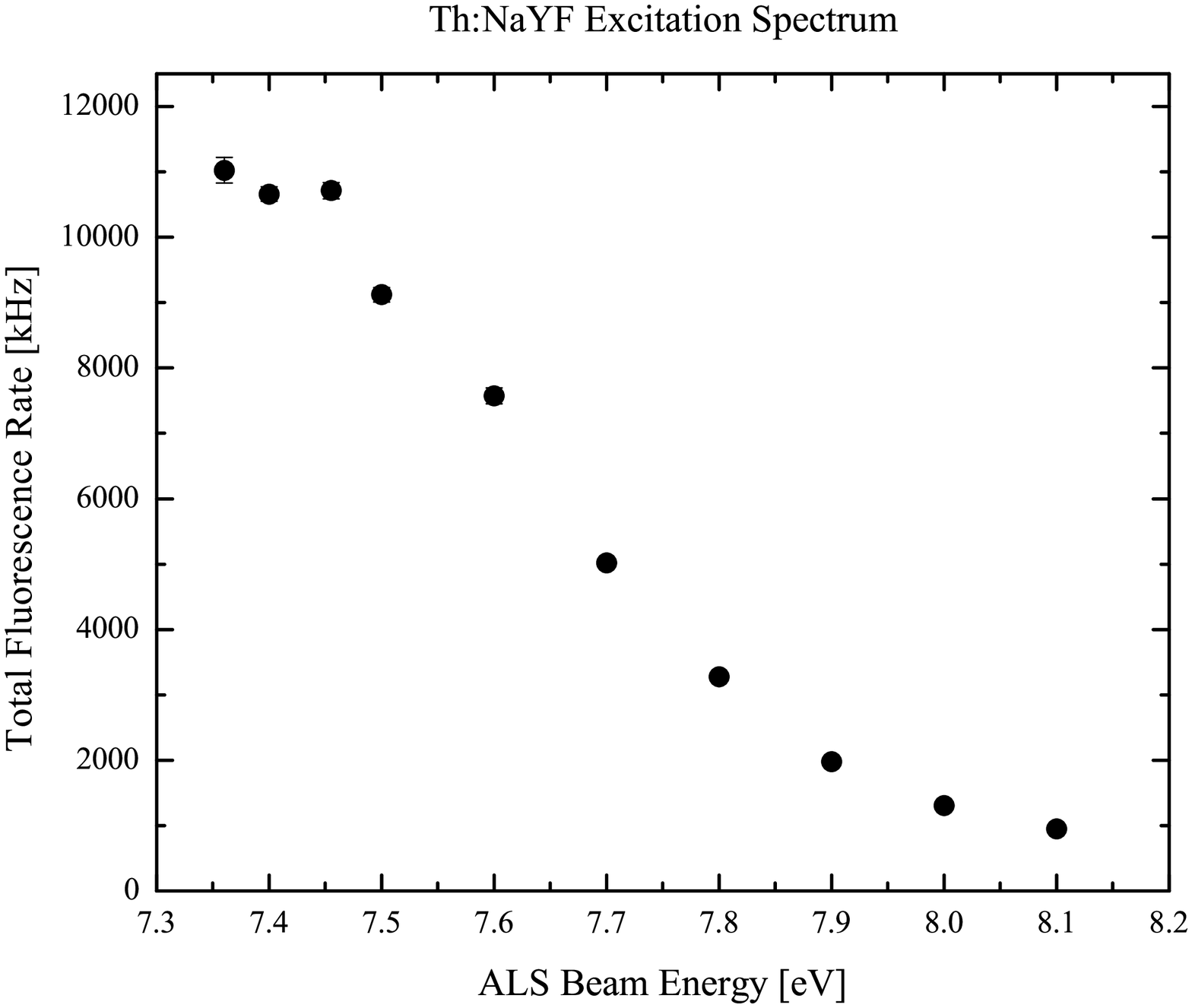}
\caption{\label{ThNaYF_excitation_spectrum}Excitation spectrum for Th:NaYF. Each point is the average of 5 repetitions with error bars given by their standard deviation.  The data have been corrected for the photon detection efficiency and the background level obtained by not illuminating the crystal has been subtracted.}
\end{minipage}\hspace{1pc}%
\begin{minipage}{19pc}
\includegraphics[width=21pc]{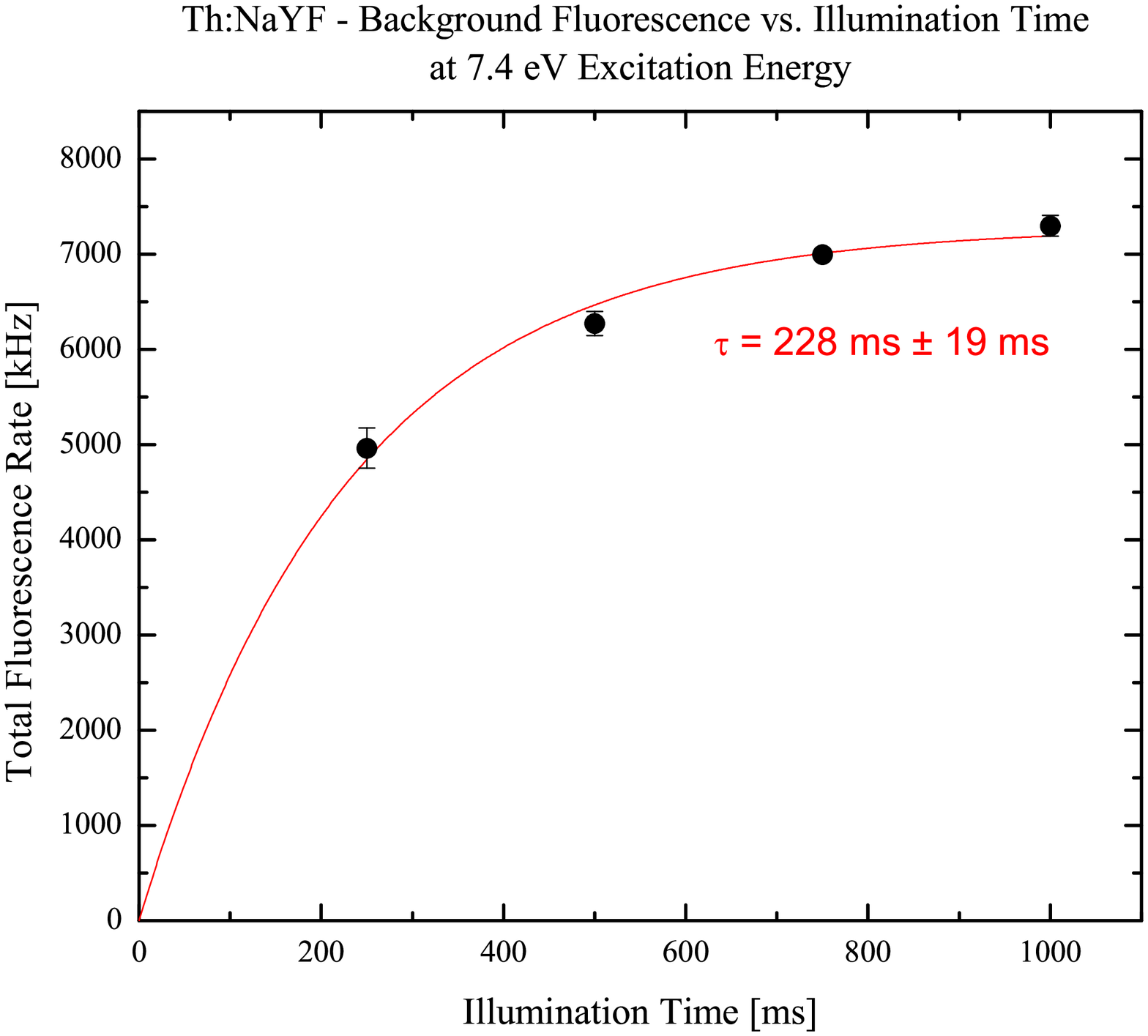}
\caption{\label{ThNaYF_signal_vs_illumination}Total fluorescence rate of Th:NaYF versus illumination time using an excitation energy of 7.4~eV.  Due to the finite closing speed of the pneumatic shutter used to block the ALS beam, the lifetime of the saturation can only be taken as an upper limit.\linebreak}
\end{minipage} 
\end{figure}

The data for Th:NaYF is shown in Fig.~\ref{ThNaYF_excitation_spectrum}.  The excitation spectrum was obtained by illuminating the crystal for 5~s with the ALS beam at a fixed energy and then collecting photons from the crystal over the next 10.7~s using the full bandwidth of the PMT (120~nm~-~220~nm).  Due to limitations in the operation of the beam and PMT shutters, there was a delay of 28~ms between the end of the illumination period and photon collection period.  At each excitation energy, the process was repeated 5 times.  The data points and their error bars in the figure are the average and standard deviation of the results.  The data shown takes into account the photon detection efficiency and the y-axis is therefore the total average fluorescence rate from the crystal over the 10.7~s of photon collection.  The background fluorescence rate, measured by leaving the VUV light beam blocked blocked during the illumination stage, was consistent with PMT dark counts ($\sim$1~Hz) and has been subtracted in the plot.  

There is clearly a very large fluorescence peak at a beam energy  $\leq$7.4~eV, however, because the ALS was in high energy mode (electron storage ring at 1.9~GeV instead of 1.5~GeV) for this run, we were unable to tune below 7.4~eV.  Nonetheless, we measured the lifetime of fluorescence arising from the 7.4~eV excitation light to be $\sim$320~ms.

We also measured the saturation of this fluorescence signal as a function of illumination time.  Due to the finite closing speed of the pneumatic shutter used to block the ALS beam, the data of Fig.~\ref{ThNaYF_signal_vs_illumination} can only be used to place an upper limit of $\sim$250~ms on the saturation rate of the fluorescence.  This indicates that longer illumination times will not yield higher VUV-induced background rates.  Finally, we note that upon inspection of the crystal after illumination, the crystal showed some radiation damage due to the VUV exciation beam.

\subsubsection{Th:YLF}\ 

\begin{figure}[h]
\includegraphics[width=21pc]{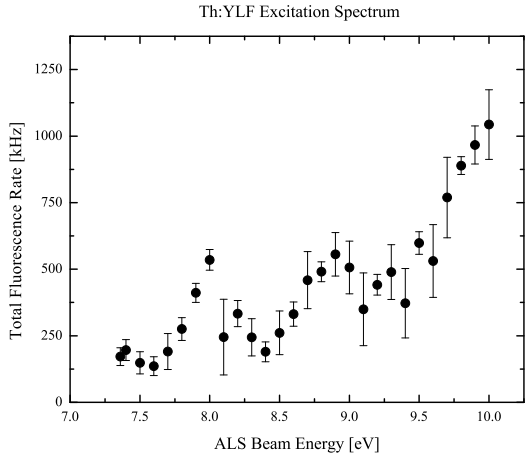}
\begin{minipage}[b]{19pc}\caption{\label{ThYLF_excitation_spectrum}Excitation spectrum for Th:YLF.}
\end{minipage} 
\end{figure}

Data for this crystal was taken in exactly the same way as the Th:NaYF crystal, and the excitation spectrum is shown in Fig.~\ref{ThYLF_excitation_spectrum}.  Here again the data points are the average of 5 repetitions at each energy, and the error bars are given by the standard deviation of those points from the average.   The fluorescence lifetimes at the peaks of 8 eV, 8.9 eV, and 10 eV were measured to be $\sim$65~ms, $\sim$70~ms and $\sim$70~ms, respectively (Table~\ref{YLFlifetimes}).  

\begin{center}
\begin{table}[h]
\caption{Lifetimes for the fluorescence peaks observed in Th:YLF.\label{YLFlifetimes}}
\centering
\begin{tabular}{@{}*{7}{l}}
  \br
  Peak & Lifetime \\
  \mr
  8.0 eV & $\sim$65~ms \\
  8.9 eV & $\sim$70~ms \\
  10.0 eV & $\sim$70~ms \\
  \br
\end{tabular}
\end{table}
\end{center}

\subsubsection{Th:LiCAF}\ 

\begin{figure}[h]
\begin{minipage}{19pc}
\includegraphics[width=21pc]{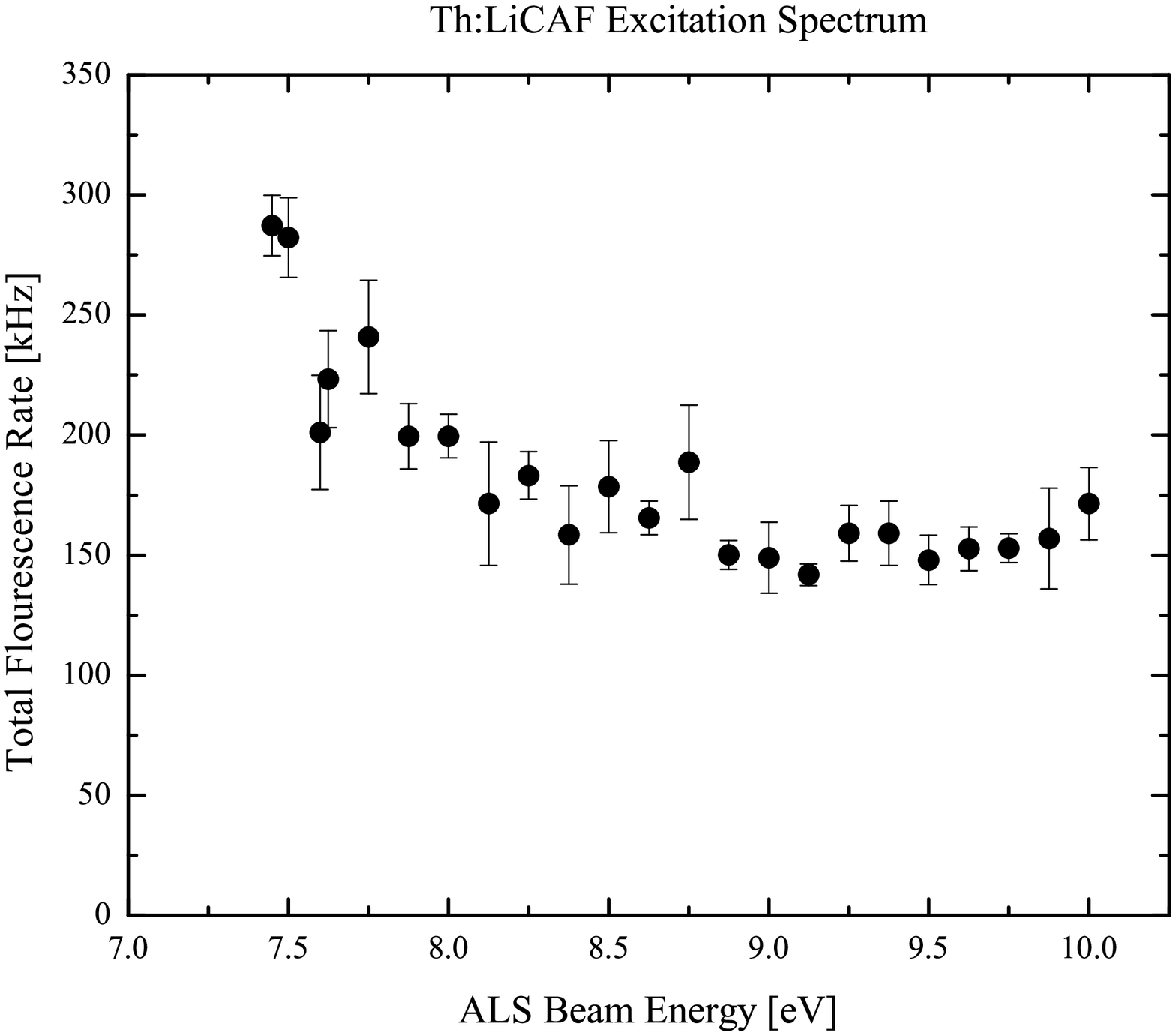}
\caption{\label{LiCAFExcitationSpectrum}Excitation spectrum of Th:LiCAF crystal.  Each data point is the average of 3 repetitions for each beam energy, and its error bar is the standard deviation of the 3 points from the average.}
\end{minipage}\hspace{1pc}%
\begin{minipage}{19pc}
\includegraphics[width=21pc]{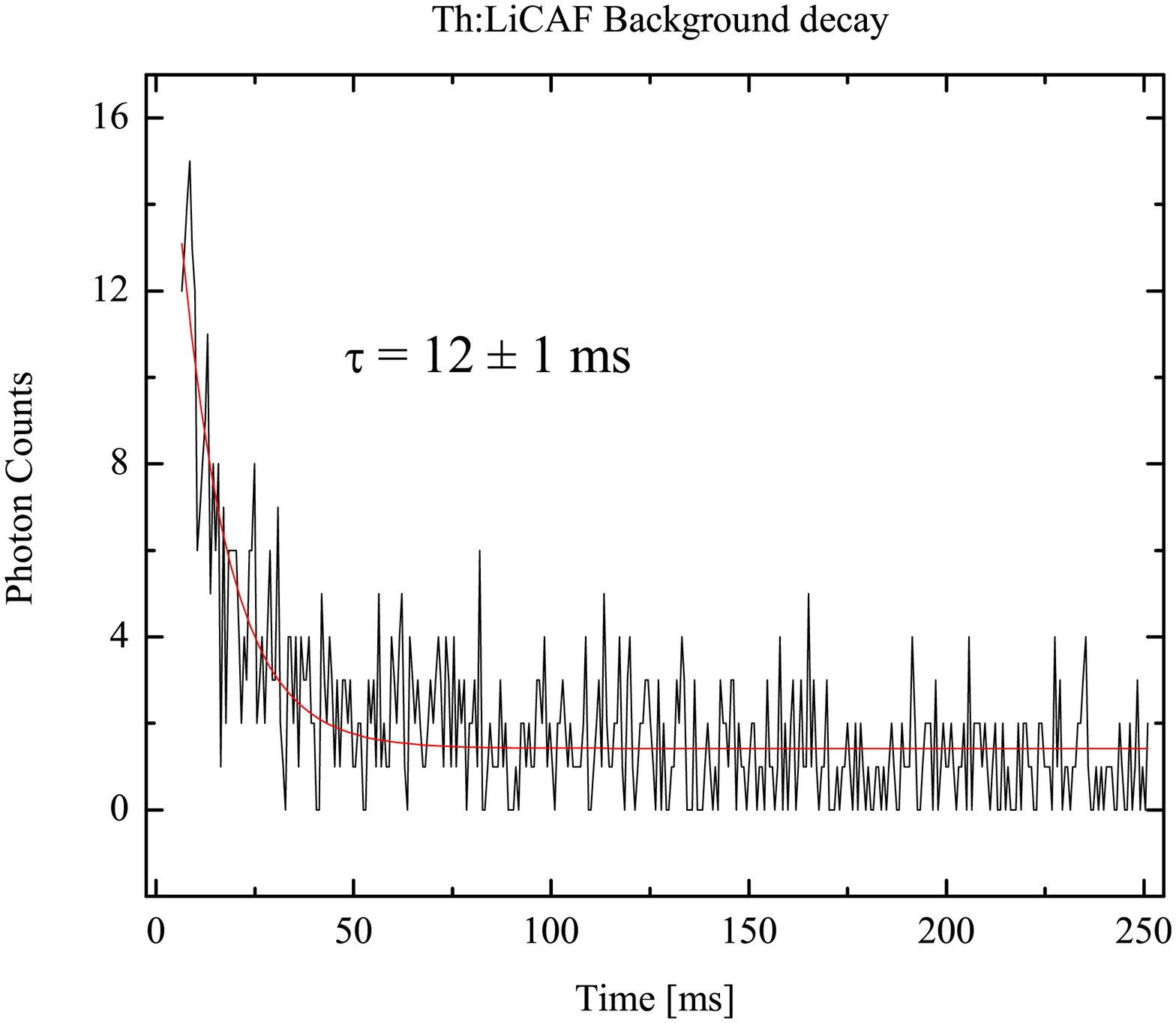}
\caption{\label{ThLiCAFBackgroundDecay}Decay of the fluorescence signal from Th:LiCAF after illuminating it at 7.4~eV.  The red line is an exponential decay fit to the data and gives 12$\pm$1~ms for the fluorescence lifetime.\linebreak }
\end{minipage} 
\end{figure}

The excitation spectrum for Th:LiCAF is shown in Fig.~\ref{LiCAFExcitationSpectrum} and was obtained in a similar way to the above crystals though the experimental details are slightly different.  Here the crystal was illuminated for 5~s, but photons were only collected for 5.4~s.  Due to improved timing controls, the delay time between illumination and photon collection was also reduced to 5~ms.  Each data point is the average of 3 repetitions for each beam energy, and its error bar is the standard deviation of the 3 points from the average.  Again, the ALS was in high energy mode for this data run, which prevented us from obtaining data below 7.4~eV.  Nevertheless, we measured the fluorescence lifetime at 7.4~eV and that data is shown in Fig.~\ref{LiCAFBackgroundDecay}.  From the fit to the data it is found that the lifetime is 12$\pm$1~ms.

 \subsubsection{Na$_2$ThF$_6$}\ 

\begin{figure}[h]
\begin{minipage}{19pc}
\includegraphics[width=21pc]{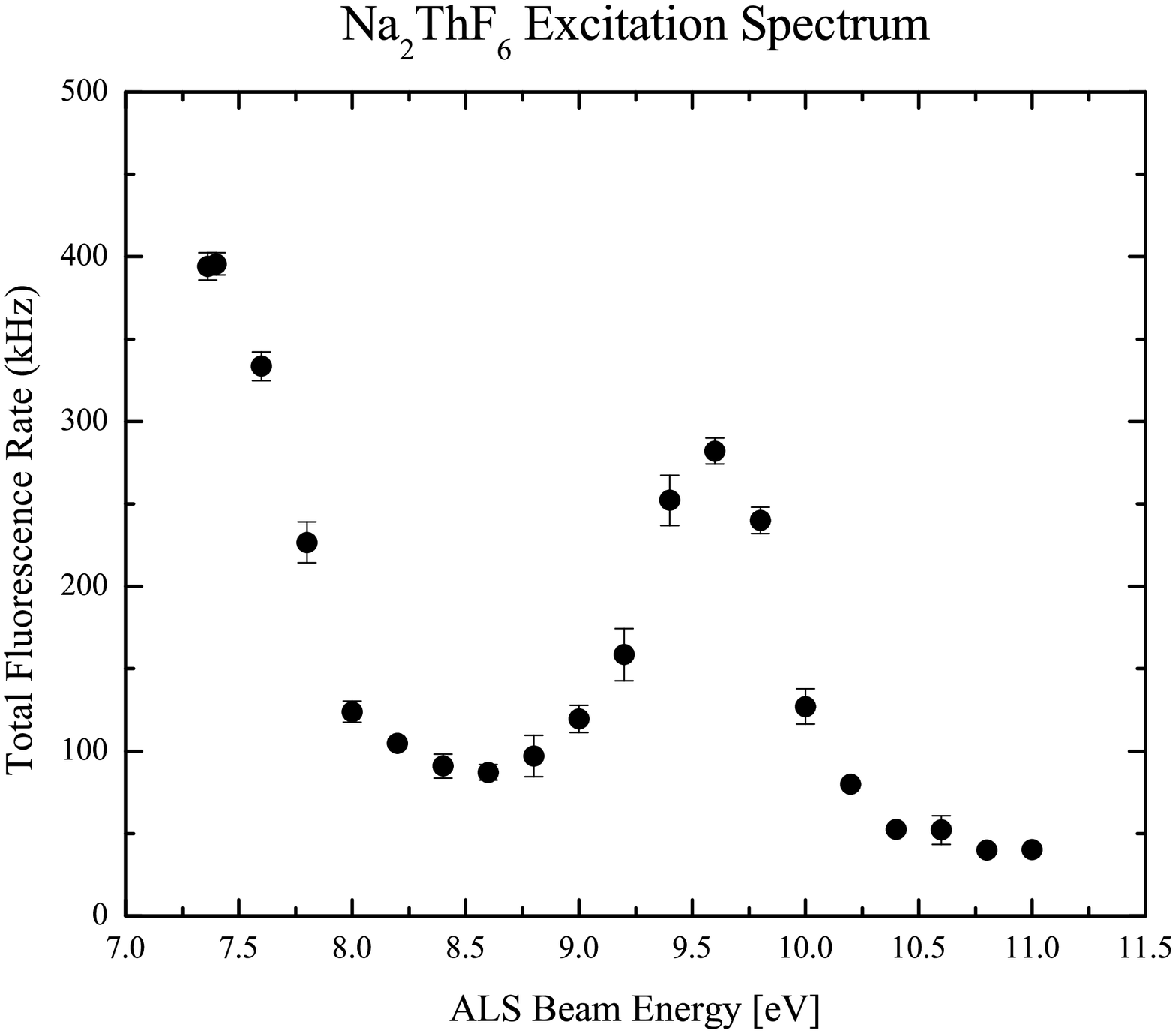}
\caption{\label{Na2ThF6ExcitationSpectrum}Excitation spectrum of Na$_2$ThF$_6$. \ \linebreak \ \linebreak \ }
\end{minipage}\hspace{1pc}%
\begin{minipage}{19pc}
\includegraphics[width=21pc]{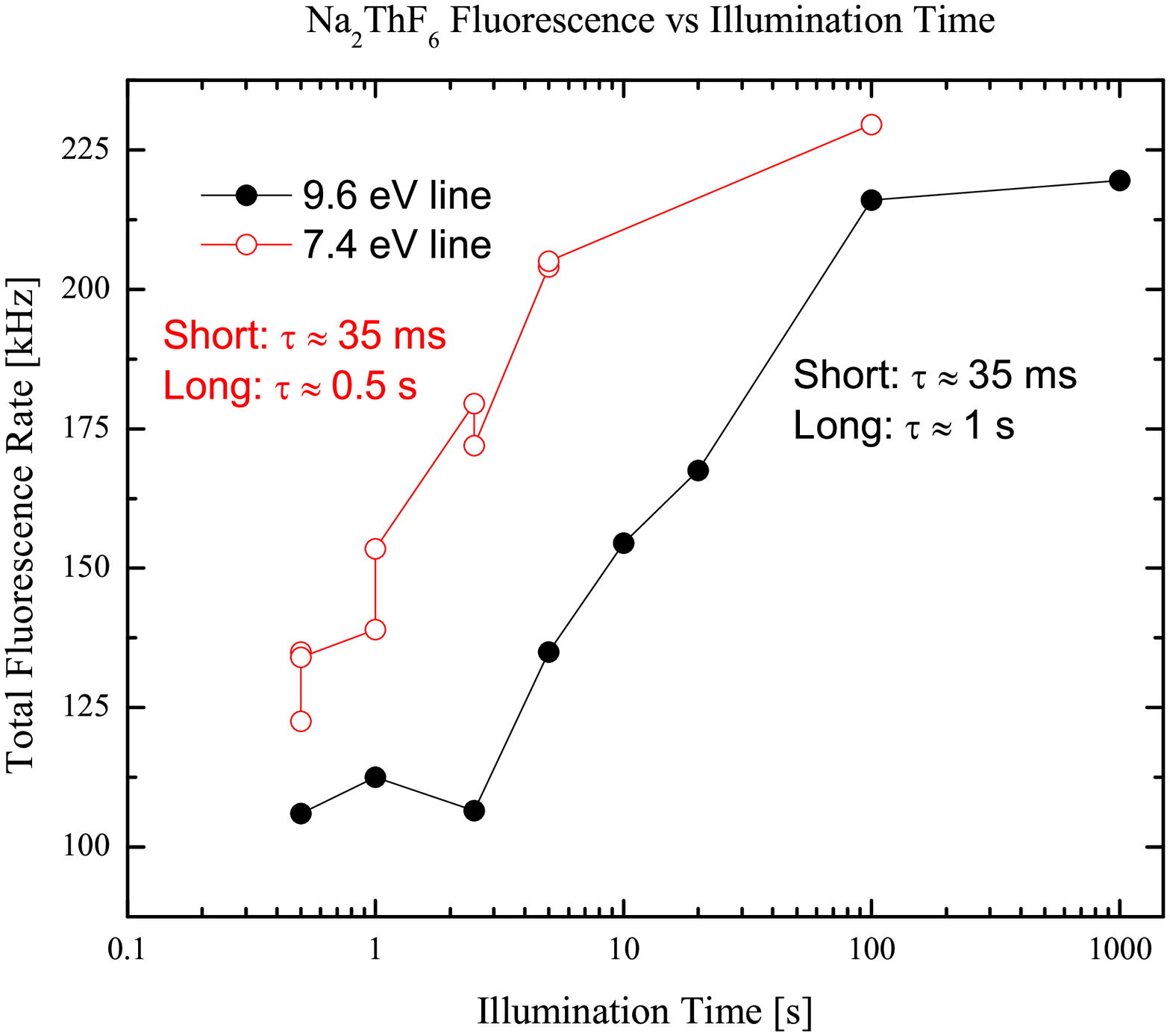}
\caption{\label{Na2ThF6_signal_vs_illumination_time}The signal versus illumination time for both fluorescence peaks observed in the Na$_2$ThF$_6$ crystal.}
\end{minipage} 
\end{figure}

This crystal is an interesting option for studies of the isomeric transition because thorium atoms are a primary constituent and thus questions about doping efficiency are removed.  Furthermore, the shifts and broadenings of the transition in the crystal due to imperfections of the doping process may be reduced.  Data for this crystal were taken in the same manner as the previous crystal and the results are shown in Fig.~\ref{Na2ThF6ExcitationSpectrum}.  The fluorescence peaks observed at an excitation envery of 7.4~eV and 9.6~eV, have both a short and a long timescale component as summarized in Table~\ref{LifetimesTable}.  The signal versus illumination time for each peak was also measured and is shown in Fig.~\ref{Na2ThF6_signal_vs_illumination_time}.  

Although the fluorescence rates are not large enough to pose a concern for using this as a host crystal in these studies, this crystal also exhibited some radiation damage due to the VUV excitation light.

\begin{center}
\begin{table}[h]
\caption{Lifetimes for the fluorescence peaks observed in Na$_2$ThF$_6$.  Both peaks displayed a short-lived ($\tau_1$) and long-lived ($\tau_2$) fluorescence component.\label{LifetimesTable}}
\centering
\begin{tabular}{@{}*{7}{l}}
  \br
  Peak & $\tau_1$  & $\tau_2$ \\
  \mr
  7.4~eV & $\sim$35~ms & $\sim$0.5~s \\
  9.6~eV & $\sim$35~ms & $\sim$1~s \\
  \br
\end{tabular}
\end{table}
\end{center}

\subsubsection{LiSAF}\ 

\begin{figure}[h]
\includegraphics[width=21pc]{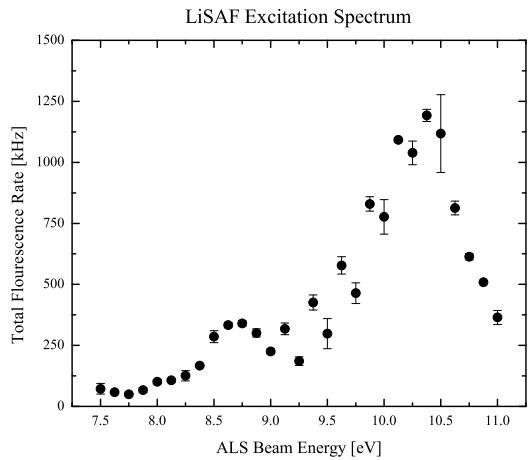}
\begin{minipage}[b]{19pc}\caption{\label{LISAF_excitation_spectrum}Excitation spectrum of an undoped LiSAF crystal.}
\end{minipage} 
\end{figure}

The data obtained for this crystal was identical to the runs performed on the Th:LiCAF and Na$_2$ThF$_6$ crystals in the previous sections.  The excitation spectrum is shown in Fig.~\ref{LISAF_excitation_spectrum}.  Near the expected isomeric transition energy, this crystal exhibits very little fluorescence, and we therefore have plans to explore it further by growing a $^{232}$Th doped version.  The largest peak observed occurs at 10.5~eV and the lifetime of the fluorescence was measured to be $\sim$500~ms. 

\subsection{Subsequent measurements of Th:LiCAF}

Based on the synchrotron data we have obtained so far, we have decided to focus on LiCAF as a host crystal for future studies using $^{229}$Th.  In addition to its low VUV-induced background fluorescence, LiCAF has several other important properties.  It is transparent down to 110~nm and has all electron spins paired. Additionally, $^{229}$Th$^{4+}$ ions are expected to substitute at the Ca$^{2+}$ site in LiCAF along with two F$^-$ interstitials for charge compensation~\cite{jackson_computer_2009}. The octahedral symmetry of the site minimizes electric field gradient effects. 

\begin{figure}[h]
\includegraphics[width=21pc]{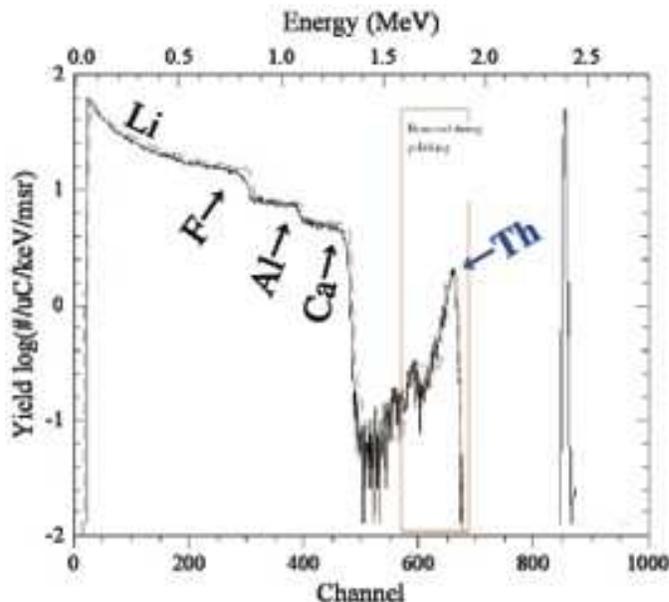}
\begin{minipage}[b]{19pc}\caption{\label{RBS_results}Results of Rutherford Backscattering measurements of the Th:LiCAF crystal for an unpolished portion of material cut out of the bulk.  The red boxed region indicates the amount of material that is removed in polishing.}
\end{minipage} 
\end{figure}

We have also measured a density in excess of 10$^{18}$ $^{232}$Th/cm$^3$ in our Th:LiCAF control crystal using the technique of Rutherford Backscattering (RBS) to interrogate the first $\approx$20~$\mu$m of the crystal.  The results of these measurements are shown in Fig.~\ref{RBS_results}. The RBS spectra shown are for an unpolished facet that we assume was cut from the bulk (\~textit{i.e.} the facet was not the top of the boule).  This facet has a thicker region (modeled as two layers) of higher concentration of Th doping. The first layer has a Th concentration of 1.8\% that is 250nm thick. The layer below that has a Th concentration of 0.3\% that is ~650nm thick. The bulk of the crystal has a concentration of $\sim$0.07\%~Th which is similiar to the polished facet (0.05\%). The Th doping of the polished facet showed a similar inhomogeneity of Th doping.

We have also performed measurements of the fluorescence backgrounds which might arise from the $\alpha$-decay of $^{229}$Th (4.8\,MeV, $\tau$~=~7880\,yr) by bombarding this crystal with a 30~nA, 4.8~MeV beam of $\alpha$-particles at the Ion Beam Materials Laboratory at Los Alamos National Laboratory.  A total fluorescence rate corresponding to 0.3~photons per $\alpha$-particle was observed over the unfiltered PMT bandwidth (Hamamatsu R7639).  Thus, a crystal containing 10$^{18}$~$^{229}$Th nulcei yields a background fluorescence rate of 2~MHz which, given the fluorescence rates observed at the ALS, is expected to be the dominant background.

\subsubsection{ALS Th:LiCAF test run}\ 

\begin{figure}[h]
\begin{minipage}{19pc}
\includegraphics[width=21pc]{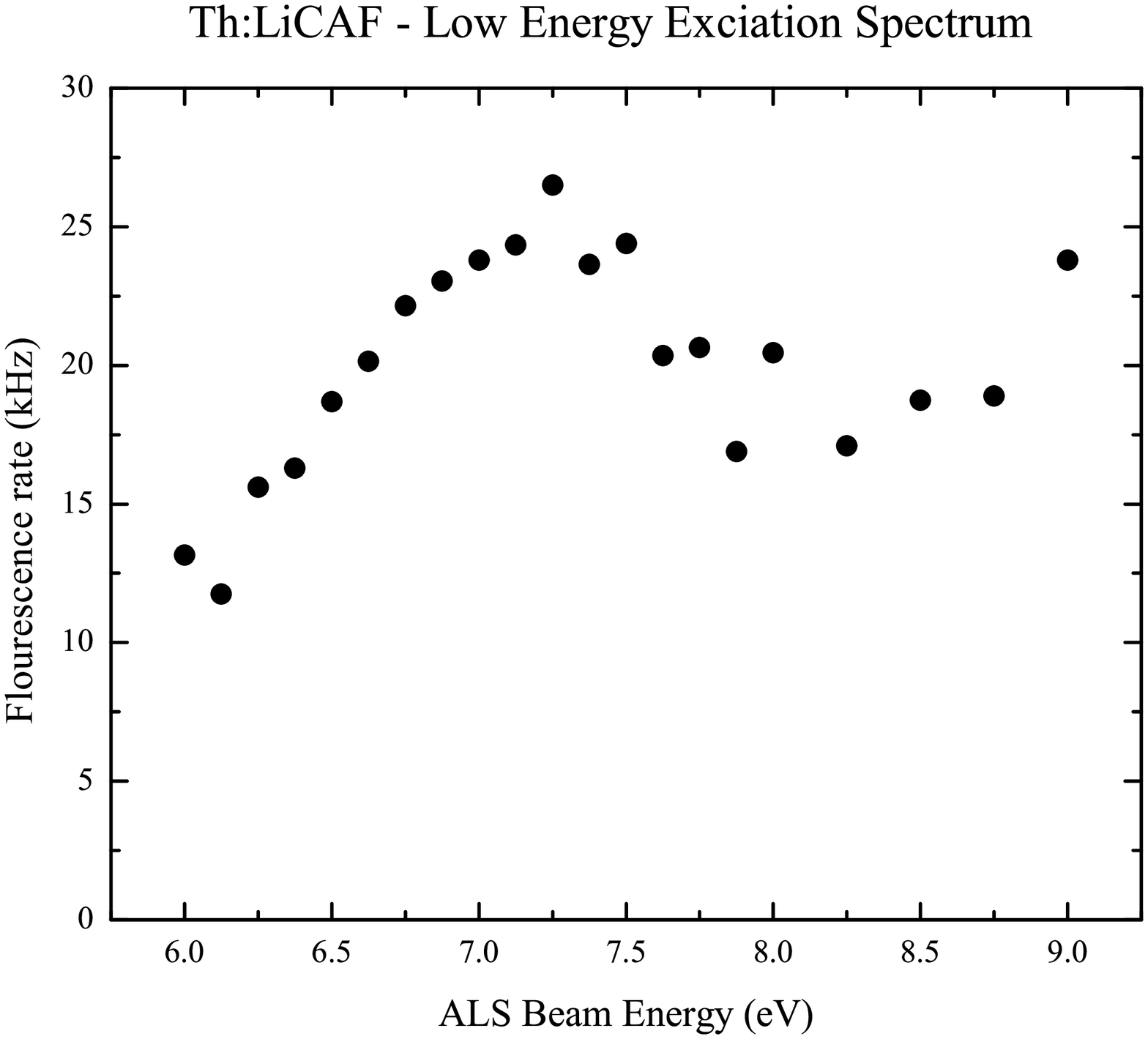}
\caption{\label{ThLiCAF_low_energy_excitation_spectrum}Excitation spectrum of the Th:LiCAF crystal across the wider beam energy range allowed by the low energy mode at the ALS.  Data is collected for $100$~s after $200$~s of illumination, just as in the proposed experimental search. }
\end{minipage}\hspace{1pc}%
\begin{minipage}{19pc}
\includegraphics[width=21pc]{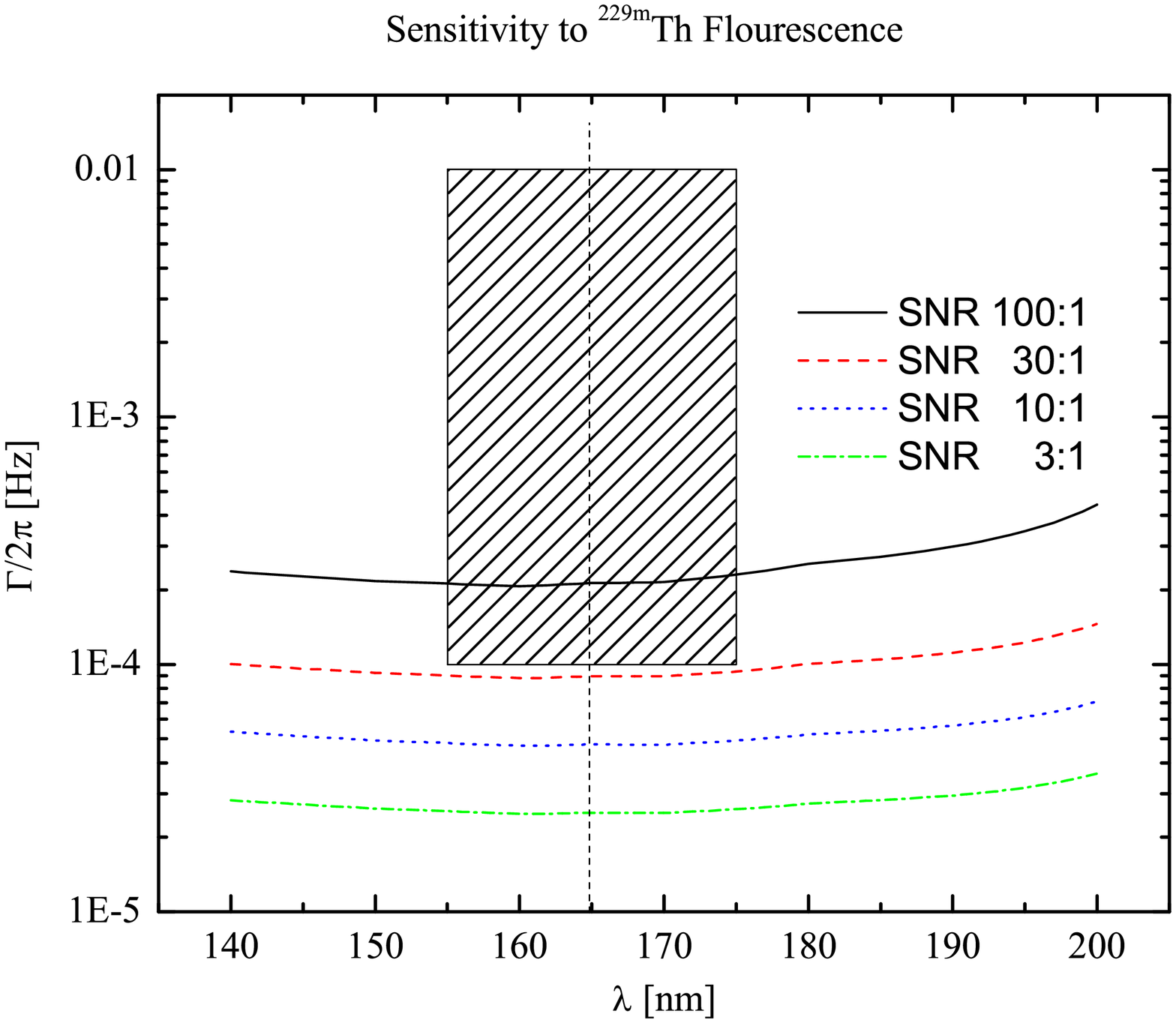}
\caption{\label{Sensitivity}Sensitivity to fluorescence from the $^{229m}$Th isomeric transition.  The shaded box indicates the region of interest for the search given the current measurement uncertainty and the expected range for the isomer lifetime. \linebreak \linebreak }
\end{minipage} 
\end{figure}

Finally, we were able to perform a separate run at the ALS when the storage ring was operated in the low energy 1.5~GeV mode.  As a result, we were able to measure the crystal using the exact experimental protocol we plan to use for measuring the isomeric transition.  The excitation spectrum, which was extended down to 6.0~eV, is shown in Fig.~\ref{ThLiCAF_low_energy_excitation_spectrum}.  Here the illumination time was 200~s, the photon collection time was 100~s, and the delay between the two was 5~ms.  Only one data point was taken for each energy.

With this data, the fluorescence rate measured in our $\alpha$-particle bombardment measurements, and the expected fluorescence rate from the isomeric transition at the ALS, we have calculated the ability of our setup to measure the isomeric transition energy depending on the natural lifetime of the transition.  The results are plotted in Fig.~\ref{Sensitivity}.  This graph shows that if the predicted range of 10-0.1~mHz is correct, we should be able to measure the transition with a SNR~$>$~30:1 during a single 8 hour shift at the ALS.

\section{Summary}

To summarize, we have characterized VUV-induced fluorescence backgrounds in several candidate host crystals that could be used in a direct search for the $^{229}$Th isomeric transition.  While none of the crystals displayed problematic backgrounds, these measurements have shown LiCAF to have the lowest fluorescence rate and shortest fluorescence lifetime overall, as well as  a high resilence to damage by the VUV beam.  Subsequent measurements have also shown it to have a low $\alpha$ particle induced background in the VUV that might arise from the alpha decay of $^{229}$Th inside a LiCAF host crystal.  Additionally, a density in excess of 10$^{18}$ $^{232}$Th nuclei/cm$^3$ was measured in our $^{232}$Th doped LiCAF crystal using RBS.  Thus, LiCAF appears to be a viable host crystal for $^{229}$Th and a direct search for the isomeric transition. 

\ack{The ALS is supported by the U.S. DOE under Contract No. $DE-AC02�05CH11231$. This work is supported by UCLRP Grant No. 09-LR-04-120497-HUDE.}

\end{document}